\documentstyle[prd,preprint,aps,epsfig,floats]{revtex}

\begin{document}
\input epsf.tex
\def\DESepsf(#1 width #2){\epsfxsize=#2 \epsfbox{#1}}
\draft
\thispagestyle{empty}
\preprint{\vbox{\hbox{OSU-HEP-01-02} \hbox{UDHEP-02-01} 
\hbox{June 2001}}}
\title{\Large \bf Classification of Effective Neutrino Mass
Operators}

\author{\large\bf K.S. Babu$^{(a)}$\footnote{E--mail:
babu@osuunx.ucc.okstate.edu} and C.N. Leung$^{(b)}$\footnote{
E--mail: leung@physics.udel.edu}}
\address{(a) Department of Physics, Oklahoma State University\\
Stillwater, OK 74078, USA \\}

\address{(b) Department of Physics and Astronomy,
University of Delaware\\
Newark, DE 19716, USA \\}
\maketitle

\thispagestyle{empty}

\begin{abstract}
We present a classification of $SU(3)_C \times SU(2)_L \times 
U(1)_Y$ gauge invariant $\Delta {\rm L} = 2$ (L being lepton 
number) effective operators relevant for generating small 
Majorana neutrino masses.  Operators of dimension up to 11 
have been included in our analysis. This approach enables us
to systematically identify interesting neutrino mass models.
It is shown that many of the well--known models fall into this 
classification.  In addition, a number of new models are 
proposed and their neutrino phenomenology is outlined.  
Of particular interest is a large class of models in which 
neutrinoless double beta decays arise at a lower order 
compared to the neutrino mass, making these decays accessible 
to the current round of experiments.  

\end{abstract}

\newpage

\section{Introduction}

Evidence for small but nonzero neutrino masses has been 
mounting over the years from experiments on atmospheric 
\cite{atmos}, solar \cite{solar} and accelerator 
\cite{LSND} neutrinos.  All experiments point toward 
a spectrum where the neutrino masses are less than or of
order 1 eV.  This is the first experimental challenge to 
the Standard Model, for its structure and particle content, 
taken as a stand--alone renormalizable field theory, do 
not allow for nonzero neutrino masses.  New ingredients must 
be introduced in order to accommodate massive neutrinos; 
and the race is on to find the underlying new physics.  

While it is possible to accommodate neutrino masses, $m_\nu$, 
by modifying the low--energy particle content of the Standard 
Model, the smallness of $m_\nu$ will remain a puzzle in such 
attempts.  Consider, for example, introducing an $SU(2)_L$ 
triplet scalar field which acquires a vacuum expectation value 
(VEV) and generates Majorana masses for the left--handed 
neutrinos \cite{cheng}.  Either the VEV of the triplet or the 
relevant Yukawa couplings must be exceedingly small in order 
to generate $m_\nu$ of order 1 eV.  Similarly, if right--handed 
neutrinos are introduced into the low energy spectrum so that 
Dirac masses can be generated, the smallness of the relevant 
Yukawa couplings ($\sim 10^{-11}$) would beg for an explanation.  

A more natural explanation for the smallness of $m_\nu$ is 
that they are generated (via some underlying new physics) 
at a scale $\Lambda$ higher than the electroweak scale\footnote{
Typically, $\Lambda$ corresponds to the scale at which lepton 
number conservation is violated.} and manifest themselves 
at low energies through effective higher dimensional ($d > 4$) 
operators which are suppressed by appropriate powers of 
$\Lambda$.  One can then understand on purely dimensional 
ground why the $m_\nu$ are small, without precise knowledge of
the underlying new physics.

As an example of this effective operator description, consider 
the well--known seesaw mechanism \cite{seesaw}.  Here the 
underlying theory has a modified particle content, viz., the 
addition of heavy right--handed neutrinos.  Upon integrating 
out these heavy fields, one arrives at an effective theory 
without the right--handed neutrinos, but with a set of 
dimension 5 operators \cite{weinberg} which can generate small 
Majorana masses for the left--handed neutrinos.   

While the $d=5$ seesaw operators are the lowest dimensional 
effective neutrino mass operators, there may be situations 
in which they are not the suitable ones.  For instance, as 
exemplified in many of the models discussed in Sec. III, 
there may be selection rules which forbid their presence.  
Such selection rules may be necessary for models in which the 
lepton number (L) breaking scale $\Lambda$ is sufficiently 
low\footnote{For example, if L is broken due to quantum 
effects of gravity, $\Lambda$ will be of order the Planck 
scale which can be as low as a few TeV in some of the 
scenarios that speculate the existence of large extra 
dimensions.} that the $d=5$ operators would generate too 
large a neutrino mass (larger than O(1) eV).  In such cases, 
the dominant contributions to the neutrino mass will come 
from operators with dimension higher than 5, typically 
through radiative corrections.  In view of the current 
interests in neutrino mass models, it will be useful to 
identify all potentially relevant neutrino mass operators.  
We list in this paper all such effective operators and 
estimate the size of the neutrino mass they generate.  In so 
doing we will reproduce many of the neutrino mass models that 
already exist in the literature and find a systematic way to 
arrive at new models.  It should be noted that this effective 
operator approach has been widely used in other studies of 
possible new physics \cite{leung}, in particular, baryon 
number violation and proton decays \cite{weinberg,wilczek}.  

Since we are interested in operators that can lead to a mass
term for the left--handed neutrino fields in the Standard 
Model, the operators must violate lepton number by two units, 
i.e., $\Delta {\rm L} = 2$.  They contain only fields which 
are present in the Standard Model and must be $SU(3)_C \times 
SU(2)_L \times U(1)_Y$ invariant.  They are also required to 
conserve baryon number, otherwise the nonobservation of
proton decays will constrain the scale $\Lambda$ to be 
greater than $10^{14}$ GeV, in which case neutrino masses 
induced by operators with $d>5$ will be too small to be 
of interest.  Because limits on lepton flavor violating 
processes such as $\mu \rightarrow eee$ and 
$\mu \rightarrow e \gamma$ constrain $\Lambda$ to be larger 
than a few TeV, we shall focus on effective operators of 
dimension 11 and lower: Operators of higher dimension will
likely lead to neutrino masses that are too small to satisfy 
the atmospheric neutrino data which require at least one 
neutrino to have a mass of about 0.03 eV \cite{atmos}.

The remainder of this paper is organized as follows.  Sec. II 
provides a classification of $\Delta {\rm L} =2$ operators 
with dimension less than 12.  In Sec. III, we present various 
renormalizable models that induce the $\Delta {\rm L} = 2$ 
operators of Sec. II.  There we also discuss the radiative 
neutrino mass generation mechanisms in these models and 
outline their main phenomenological consequences.  In Sec. IV, 
we point out the significance of some of the operators to 
neutrinoless double beta ($\beta \beta 0 \nu$) decays.  We 
identify a number of effective operators which induce 
$\beta \beta 0 \nu$ decays at a lower order compared to the 
neutrino mass.  We offer our conclusions in Sec. V.

\section{Classification of effective $\Delta {\rm L} = 2$ 
operators}

We shall use a notation in which all fermion fields are
left--handed, denoted by
\begin{equation}
L(1,2,-{1 \over 2}), ~ e^c(1,1,1),~ Q(3,2,{1 \over 6}),
~ d^c(\bar{3},1,{1 \over 3}),~u^c(\bar{3},1,-{2 \over 3})~,
\end{equation}
where the $SU(3)_C \times SU(2)_L \times U(1)_Y$ quantum numbers
are indicated in parentheses.  We shall suppress all generation 
indices in this section, but will reinstate them in the next 
section when we discuss renormalizable neutrino mass models.  
Here $L$ and $Q$ stand for the lepton and quark doublets, 
respectively; $e^c = C\bar{e_R}^T$ ($C$ being the charge 
conjugation operator), $u^c$, and $d^c$ stand for 
the charge conjugates of the right--handed charged leptons, 
up--type quarks, and down--type quarks, respectively.  The 
Standard Model Higgs doublet is denoted as $H(1,2,{1 \over 2})$ 
and $\bar{H}$ will denote its hermitian conjugate.

The $\Delta {\rm L} = 2$ operators can be derived systematically 
using the Standard Model degrees of freedom as follows.  There 
are three basic fermion bilinears that carry two units of 
letpon number:
\begin{equation}
\{L^i L^j,~ L^i \bar{e^c},~\bar{e^c} \bar{e^c}\}_.
\end{equation}
Here $i$ and $j$ are $SU(2)_L$ indices, and $\bar{e^c}$ 
stands for either the hermitian conjugate or the Dirac adjoint 
of $e^c$.  Any $\Delta {\rm L} = 2$ effective operator will 
have one of these basic fermion bilinears accompanied by a 
product of other fields which is neutral under color and 
carries a net baryon number of zero.  We classify the 
effective neutrino mass operators according to the number of 
fermion fields they contain.  Three separate groups can be 
identified: (i) operators containing $L^i L^j$ and no other 
fermion fields; (ii) operators containing four fermion 
fields; and (iii) operators containing six fermion fields.  
Operators containing four or more fermion bilinears have 
dimension 12 or higher and will not be considered here because, 
as mentioned above, the neutrino masses generated by such 
operators will be constrained by limits on lepton flavor 
violation to be typically much smaller than $0.03$ eV and will 
not be that interesting for the current neutrino oscillation 
phenomenology.  In case (i), neutrino masses will arise at 
tree level.  In case (ii), one pair of fermion fields must be 
annihilated to generate neutrino masses, which will therefore 
arise at the one--loop level.  And in case (iii), which 
requires the annihilation of two fermion pairs, neutrino 
masses will arise as two--loop radiative corrections.

(i) With $L^i L^j$ not accompanied by any more fermion fields, 
one obtains the well--known dimension five operator for 
neutrino mass \cite{weinberg}:
\begin{equation}
{\cal O}_1 = L^i L^j H^k H^l \epsilon_{ik} \epsilon_{jl}
\label{O1}
\end{equation}

(ii) Operators with four fermion fields are:

\begin{eqnarray}
{\cal O}_2&=& L^i L^j L^k e^c H^l \epsilon_{ij} \epsilon_{kl}
\nonumber \\
{\cal O}_3&=& \{L^i L^j Q^k d^c H^l \epsilon_{ij} \epsilon_{kl},
~~L^i L^j Q^k d^c H^l \epsilon_{ik} \epsilon_{jl}\}  \nonumber \\
{\cal O}_4&=& \{L^i L^j \bar{Q}_i \bar{u^c} H^k \epsilon_{jk},~~
L^i L^j \bar{Q}_k\bar{u^c}H^k \epsilon_{ij}\}
\nonumber \\
{\cal O}_5 &=& L^i L^j Q^k d^c H^l H^m \bar{H}_i \epsilon_{jl}
\epsilon_{km} \nonumber \\
{\cal O}_6 &=& L^i L^j \bar{Q}_k\bar{u^c}H^l H^k \bar{H}_i
\epsilon_{jl} \nonumber \\
{\cal O}_{7} &=& L^iQ^j \bar{e^c}\bar{Q}_kH^k H^l H^m 
\epsilon_{il} \epsilon_{jm} \nonumber \\
{\cal O}_{8} &=& L^i \bar{e^c} \bar{u^c} d^c H^j \epsilon_{ij}
\label{O4f}
\end{eqnarray}

Before we list the operators in group (iii), we wish to make 
several remarks.

\begin{enumerate}

\item Operators ${\cal O}_{2} - {\cal O}_6$ are obtained by
multiplying $L^i L^j$ with any one of the combinations
$\{ Le^c,~ Q d^c,~Q u^c,~\bar{L}\bar{e^c},~\bar{Q} \bar{d^c},
~ \bar{Q} \bar{u^c}\}$ which all conserve lepton as well as 
baryon number. ${\cal O}_{7}$ and ${\cal O}_{8}$ are 
obtained as the product of $L^i \bar{e^c}$ with $\bar{F} F$ 
where $F=\{L,e^c,Q,u^c,d^c\}$.  For operators without any 
derivatives or gauge boson fields (see comment 5 below), 
the number of barred (and unbarred) fields should be even.  
Note also that there are no group (ii) operators of the 
type $\bar{e^c} \bar{e^c}$ that are gauge invariant, owing 
to $SU(2)_L$ antisymmetry in the Higgs field.

\item We have shown explicitly the $SU(2)_L$ group contractions 
using the indices ($i$,$j$,$k$,$l$,$m$,$n$).  The color indices
are, however, suppressed.  In operators involving six fermion
fields (to follow), when there are two quark and two anti--quark
fields, there are two possible color contractions, which will
not be shown, but should be assumed.

\item Lorentz indices are not explicitly shown in our 
operators.  All possible Lorentz contractions should be
allowed.  Operator ${\cal O}_1$ is a unique Lorentz scalar,
viz., $(L^{iT} C L^j)H^k H^l \epsilon_{ik} \epsilon_{jl}$,
where $C$ is the charge conjugation matrix.  Operator
${\cal O}_2$ has the following Lorentz contractions:
\begin{eqnarray}
&~& (L^{iT}C L^j)(L^{kT} Ce^c)H^l \epsilon_{ij} \epsilon_{kl},~
(L^{iT}C L^k)(L^{jT} Ce^c)H^l \epsilon_{ij} \epsilon_{kl},~
\nonumber \\
&~& (L^{iT}C e^c)(L^{jT} CL^k)H^l \epsilon_{ij} \epsilon_{kl}~,
\label{O2}
\end{eqnarray}
along with those obtained from Eq. (\ref{O2}) by replacing $C$ 
with $\sigma_{\mu \nu}$.  In the first entry for the operator
${\cal O}_4$, the following Lorentz contractions are allowed:
\begin{eqnarray}
{\cal O}_4 &=& (L^{iT} C L^j)(\bar{Q}_i^T C \bar{u^c})H^k
\epsilon_{jk},~(L^{iT} \sigma_{\mu \nu} L^j)(\bar{Q}_i^T
\sigma^{\mu \nu} \bar{u^c})H^k \epsilon_{jk},~ \nonumber \\
&~& (\bar{Q}_i \gamma_\mu L^{i})(\bar{u^c} \gamma^\mu L^j)H^k
\epsilon_{jk},~(\bar{Q}_i \gamma_\mu L^{j})(\bar{u^c}
\gamma^\mu L^i)H^k \epsilon_{jk}~.
\end{eqnarray}
Although Fierz identities exist between several of these 
operators, they must be counted as independent because 
the $SU(2)_L$ contractions will not be the same.

\item We have not explicitly written down operators with
$\Delta {\rm L} = 2$ that can be obtained by multiplying the 
lower dimensional $\Delta {\rm L} = 2$ operators with one of 
the gauge invariant, baryon and lepton number conserving 
operators that appear in the Standard Model Lagrangian.  These 
latter operators include $\bar{H}_i H^i, ~L^ie^c \bar{H}_i,
~Q^i d^c \bar{H}_i$, and $Q^i u^c H^j \epsilon_{ij}$, as well 
as their hermitian conjugates.  Such product operators, 
while not explicitly displayed, are understood to be present 
in our list.  In the case of product operators containing 
$\bar{H}_iH^i$, with the Higgs doublets having the trivial 
$SU(2)_L$ contraction as shown, the presence of such a 
higher dimensional operator will necessarily imply the 
existence of the corresponding lower dimensional operator 
without the $\bar{H}_i H^i$ factor.  (This can be seen by 
closing the $\bar{H}_i$ and $H^i$ lines to form a loop, 
which will give an infinite contribution to the lower 
dimensional operator, necessitating its presence.)  When 
the $SU(2)_L$ contraction on the Higgs fields is nontrivial, 
we do list explicitly such operators (compare, e.g., 
${\cal O}_3$ with ${\cal O}_5$).  Product operators involving 
one of the Standard Model operators $L^i e^c {\bar H}_i$, 
$Q^id^c {\bar H}_i$, or $Q^i u^c H^j \epsilon_{ij}$ (or their 
hermitian conjugates), and one of the lower dimensional 
$\Delta {\rm L} = 2$ operators of Eqs. (3)-(4), while not
listed explicitly,  can be quite interesting from the point 
of view of neutrino mass models.  Examples of this type of 
operators are $L^i L^j \bar{L}_k \bar{e^c} H^l H^m H^k 
\epsilon_{il} \epsilon_{jm}$, $L^i L^j \bar{Q}_k \bar{d^c} 
H^l H^m H^k \epsilon_{il} \epsilon_{jm}$, and
$L^i L^j Q^k u^c H^l H^m H^n \epsilon_{il} \epsilon_{jm}
\epsilon_{kn}$.  Since all possible Lorentz contractions are 
to be assumed, these operators can be generated without 
inducing the lower dimensional operator ${\cal O}_1$.  We 
shall give an example of this type of neutrino mass models in 
Sec. III.  Note also that the flavor structure of $L^i e^c 
{\bar H}_i$, etc., in these product operators need not be the 
same as that of the corresponding  Standard Model operators, 
which implies the possibility of interesting neutrino 
phenomenology from these operators.  

\item We have not included in our list $\Delta {\rm L} = 2$ 
operators which involve the Standard Model gauge boson fields.  
They can arise, for example, through covariant derivatives of 
the fermion or Higgs fields.  It may be more difficult 
to generate such operators at tree level from an underlying 
renormalizable theory, making them perhaps less interesting for 
the generation of neutrino masses.  Nevertheless, we wish to 
list such $\Delta {\rm L} =2$ operators of the lowest dimension, 
which turns out to be $7$.  They are $(L^T \sigma_{\mu \nu} L) 
H H B^{\mu \nu}$, $(L^T \sigma_{\mu \nu} L) H H W^{\mu \nu}$, 
$(L^T C D_\mu D^\mu L) H H$, and $(\bar{e^c} \gamma_\mu D^\mu L) 
H H H$.  Here $B^{\mu \nu}$ and $W^{\mu \nu}$ are the $U(1)_Y$ 
and $SU(2)_L$ field strength tensors and we have suppressed 
the $SU(2)_L$ indices for simplicity.  Although we have only 
shown operators with the covariant derivative acting on a 
specific field, it is understood that one should include 
similar operators with the covariant derivative acting on the 
other fields.  For example, the third operator listed above 
includes $(L^T C L) (D_\mu H) (D^\mu H)$, and the fourth 
operator includes $(\bar{e^c} \gamma_\mu L) (D^\mu H) H H$, 
and so on.

\end{enumerate}

(iii) We now proceed to write down the operators with six 
fermion fields through dimension 11.  The procedure we follow 
is analogous to the case of the operators containing four 
fermion fields. There are 12 such operators at the dimension 9 
level:

\begin{eqnarray}
{\cal O}_{9}&=&L^i L^j L^k e^c L^l e^c \epsilon_{ij}
\epsilon_{kl} \nonumber \\
{\cal O}_{10}&=&L^i L^j L^k e^c Q^l d^c \epsilon_{ij}
\epsilon_{kl} \nonumber \\
{\cal O}_{11}&=&\{L^i L^j Q^k d^c Q^l d^c \epsilon_{ij}
\epsilon_{kl},~~L^i L^j Q^k d^c Q^l d^c \epsilon_{ik}
\epsilon_{jl}\} \nonumber \\
{\cal O}_{12}&=&\{L^i L^j \bar{Q}_i \bar{u^c} \bar{Q_j}
\bar{u^c},~~L^i L^j \bar{Q}_k \bar{u^c} \bar{Q}_l \bar{u^c}
\epsilon_{ij} \epsilon^{kl} \}
\nonumber \\
{\cal O}_{13}&=&L^i L^j \bar{Q}_i \bar{u^c}L^l e^c
\epsilon_{jl} \nonumber \\
{\cal O}_{14}&=&\{L^i L^j \bar{Q}_k \bar{u^c} Q^k d^c
\epsilon_{ij},~~L^i L^j \bar{Q}_i \bar{u^c} Q^l d^c
\epsilon_{jl}\} \nonumber \\
{\cal O}_{15} &=& L^i L^j L^k d^c \bar{L}_i \bar{u^c}
\epsilon_{jk} \nonumber \\
{\cal O}_{16} &=& L^i L^j e^c d^c \bar{e^c} \bar{u^c}
\epsilon_{ij} \nonumber \\
{\cal O}_{17}&=& L^i L^j d^c d^c \bar{d^c} \bar{u^c}
\epsilon_{ij} \nonumber \\
{\cal O}_{18}&=& L^i L^j d^c u^c \bar{u^c} \bar{u^c}
\epsilon_{ij} \nonumber \\
{\cal O}_{19}&=& L^i Q^j d^c d^c \bar{e^c} \bar{u^c}
\epsilon_{ij} \nonumber \\
{\cal O}_{20} &=& L^i d^c \bar{Q}_i \bar{u^c} \bar{e^c}
\bar{u^c}
\label{O6fd9}
\end{eqnarray}

And there are 40 operators with $d=11$: 

\begin{eqnarray}
{\cal O}_{21}&=& \{L^i L^j L^k e^c Q^l u^c H^m H^n
\epsilon_{ij} \epsilon_{km} \epsilon_{ln},~~~
L^i L^j L^k e^c Q^l u^c H^m H^n
\epsilon_{il} \epsilon_{jm} \epsilon_{kn}\}
 \nonumber \\
{\cal O}_{22} &=& 
L^i L^j L^k e^c \bar{L}_k \bar{e^c} H^l H^m \epsilon_{il}
\epsilon_{jm} \nonumber \\
{\cal O}_{23} &=& L^i L^jL^k e^c \bar{Q}_k\bar{d^c}H^l
H^m \epsilon_{il} \epsilon_{jm} \nonumber \\
{\cal O}_{24} &=& \{L^i L^j Q^k d^c Q^l d^c H^m \bar{H}_i
\epsilon_{jk} \epsilon_{lm},~~~ L^i L^j Q^k d^c Q^l d^c
H^m \bar{H}_i \epsilon_{jm} \epsilon_{kl}\} \nonumber \\
{\cal O}_{25} &=&  
L^i L^j Q^k d^c Q^l u^c H^m H^n \epsilon_{im}
\epsilon_{jn} \epsilon_{kl} \nonumber \\
{\cal O}_{26} &=& \{L^i L^j Q^k d^c \bar{L}_i \bar{e^c}
H^l H^m \epsilon_{jl} \epsilon_{km},~~~L^i L^j Q^k d^c
\bar{L}_k \bar{e^c} H^l H^m \epsilon_{il} \epsilon_{jm}\}
\nonumber \\
{\cal O}_{27} &=& \{L^i L^j Q^k d^c \bar{Q}_i\bar{d^c} H^l
H^m \epsilon_{jl} \epsilon_{km},~~~L^i L^j Q^k d^c
\bar{Q}_k\bar{d^c} H^l H^m \epsilon_{il} \epsilon_{jm}\}
\nonumber \\
{\cal O}_{28} &=& \{L^i L^j Q^k d^c \bar{Q}_j \bar{u^c}
H^l \bar{H}_i \epsilon_{kl},~~~L^i L^j Q^k d^c \bar{Q}_k
\bar{u^c} H^l \bar{H}_i \epsilon_{jl}, \nonumber \\
&~& L^i L^j Q^k d^c \bar{Q}_l \bar{u^c} H^l \bar{H}_i
\epsilon_{jk}\} \nonumber \\
{\cal O}_{29} &=& \{L^i L^j Q^k u^c \bar{Q}_k \bar{u^c} 
H^l H^m \epsilon_{il} \epsilon_{jm},~~~L^i L^j Q^k u^c
\bar{Q}_l \bar{u^c} H^l H^m \epsilon_{ik} \epsilon_{jm}\}
\nonumber \\
{\cal O}_{30} &=& \{L^i L^j \bar{L}_i \bar{e^c}
\bar{Q}_k\bar{u^c} H^k H^l \epsilon_{jl},~~~L^i L^j
\bar{L}_m \bar{e^c} \bar{Q}_n \bar{u^c} H^k H^l
\epsilon_{ik} \epsilon_{jl} \epsilon^{mn}\} \nonumber \\
{\cal O}_{31} &=& \{L^i L^j \bar{Q}_i\bar{d^c}
\bar{Q}_k\bar{u^c} H^k H^l \epsilon_{jl},~~~L^i L^j
\bar{Q}_m\bar{d^c} \bar{Q}_n\bar{u^c} H^k H^l
\epsilon_{ik} \epsilon_{jl} \epsilon^{mn}\} \nonumber \\
{\cal O}_{32} &=& \{L^i L^j \bar{Q}_j \bar{u^c}
\bar{Q}_k \bar{u^c} H^k \bar{H}_i,~~~L^i L^j
\bar{Q}_m \bar{u^c} \bar{Q}_n \bar{u^c} H^k
\bar{H}_i \epsilon_{jk} \epsilon^{mn}\} \nonumber \\
{\cal O}_{33} &=& \bar{e^c} \bar{e^c} L^i L^j e^c e^c H^k
H^l \epsilon_{ik} \epsilon_{jl} \nonumber \\
{\cal O}_{34} &=& \bar{e^c} \bar{e^c} L^i Q^j e^c d^c H^k
H^l \epsilon_{ik} \epsilon_{jl} \nonumber \\
{\cal O}_{35} &=& \bar{e^c} \bar{e^c} L^i e^c \bar{Q}_j
\bar{u^c} H^j H^k \epsilon_{ik} \nonumber \\
{\cal O}_{36} &=& \bar{e^c} \bar{e^c} Q^i d^c Q^j d^c H^k
H^l \epsilon_{ik} \epsilon_{jl} \nonumber \\
{\cal O}_{37} &=& \bar{e^c} \bar{e^c} Q^i d^c \bar{Q}_j
\bar{u^c} H^j H^k \epsilon_{ik} \nonumber \\
{\cal O}_{38} &=& \bar{e^c} \bar{e^c} \bar{Q}_i \bar{u^c}
\bar{Q}_j \bar{u^c} H^i H^j \nonumber \\
{\cal O}_{39} &=& \{L^i L^j L^k L^l \bar{L}_i \bar{L}_j
H^m H^n \epsilon_{jm} \epsilon_{kl},~~~L^i L^j L^k L^l
\bar{L}_m \bar{L}_n H^m H^n \epsilon_{ij} \epsilon_{kl},
\nonumber \\
&~& L^i L^j L^k L^l \bar{L}_i \bar{L}_m H^m H^n
\epsilon_{jk} \epsilon_{ln},~~~~L^i L^j L^k L^l \bar{L}_p
\bar{L}_q H^m H^n \epsilon_{ij} \epsilon_{km} \epsilon_{ln}
\epsilon^{pq}\} \nonumber \\
{\cal O}_{40} &=& \{L^i L^j L^k Q^l \bar{L}_i \bar{Q}_j
H^m H^n \epsilon_{km} \epsilon_{ln},~~~L^i L^j L^k Q^l
\bar{L}_i \bar{Q}_l H^m H^n \epsilon_{jm} \epsilon_{kn},
\nonumber \\
&~& L^i L^j L^k Q^l \bar{L}_l \bar{Q}_i H^m H^n
\epsilon_{jm} \epsilon_{kn}, ~~~~L^i L^j L^k Q^l \bar{L}_i
\bar{Q}_m H^m H^n \epsilon_{jk} \epsilon_{ln}, \nonumber \\
&~& L^i L^j L^k Q^l \bar{L}_i \bar{Q}_m H^m H^n \epsilon_{jl}
\epsilon_{kn}, ~~~L^i L^j L^k Q^l \bar{L}_m \bar{Q}_i H^m
H^n \epsilon_{jk} \epsilon_{ln}, \nonumber \\
&~& L^i L^j L^k Q^l \bar{L}_m \bar{Q}_i H^m H^n \epsilon_{jl}
\epsilon_{kn},~~~~L^i L^j L^k Q^l \bar{L}_m \bar{Q}_n H^m H^n
\epsilon_{ij} \epsilon_{kl}, \nonumber \\
&~& L^i L^j L^k Q^l \bar{L}_m \bar{Q}_n H^p H^q \epsilon_{ip}
\epsilon_{jq} \epsilon_{kl} \epsilon^{mn},~~~
L^i L^j L^k Q^l \bar{L}_m \bar{Q}_n H^p H^q \epsilon_{ip}
\epsilon_{lq} \epsilon_{jk} \epsilon^{mn}\} \nonumber \\
{\cal O}_{41} &=& \{L^i L^j L^k d^c \bar{L}_i \bar{d^c} H^l
H^m \epsilon_{jl} \epsilon_{km},~~~L^i L^j L^k d^c \bar{L}_l
\bar{d^c} H^l H^m \epsilon_{ij} \epsilon_{km}\} \nonumber \\
{\cal O}_{42} &=& \{L^i L^j L^k u^c \bar{L}_i \bar{u^c} H^l
H^m \epsilon_{jl} \epsilon_{km},~~~L^i L^j L^k u^c \bar{L}_l
\bar{u^c} H^l H^m \epsilon_{ij} \epsilon_{km}\} \nonumber \\
{\cal O}_{43} &=& \{L^i L^j L^k d^c \bar{L}_l\bar{u^c} H^l
\bar{H}_i \epsilon_{jk},~~~L^i L^j L^k d^c \bar{L}_j\bar{u^c}
H^l \bar{H}_i \epsilon_{kl}, \nonumber \\
&~& L^i L^j L^k d^c \bar{L}_l\bar{u^c} H^m \bar{H}_n
\epsilon_{ij} \epsilon_{km} \epsilon^{ln}\} \nonumber \\
{\cal O}_{44} &=& \{L^i L^j Q^k e^c \bar{Q}_i \bar{e^c} H^l
H^m \epsilon_{jl} \epsilon_{km},~~~L^i L^j Q^k e^c \bar{Q}_k
\bar{e^c} H^l H^m \epsilon_{il} \epsilon_{jm}, \nonumber \\
&~& L^i L^j Q^k e^c \bar{Q}_l \bar{e^c} H^l H^m \epsilon_{ij}
\epsilon_{km},~~~L^i L^j Q^k e^c \bar{Q}_l \bar{e^c} H^l H^m
\epsilon_{ik} \epsilon_{jm}\} \nonumber \\
{\cal O}_{45} &=& L^i L^j e^c d^c \bar{e^c} \bar{d^c} H^k H^l
\epsilon_{ik} \epsilon_{jl} \nonumber \\
{\cal O}_{46} &=& L^i L^j e^c u^c \bar{e^c} \bar{u^c} H^k H^l
\epsilon_{ik} \epsilon_{jl} \nonumber \\
{\cal O}_{47} &=& \{L^i L^j Q^k Q^l \bar{Q}_i\bar{Q}_j H^m H^n
\epsilon_{km} \epsilon_{ln},~~~L^i L^j Q^k Q^l \bar{Q}_i
\bar{Q}_k H^m H^n \epsilon_{jm} \epsilon_{ln}, \nonumber \\
&~& L^i L^j Q^k Q^l \bar{Q}_k\bar{Q}_l H^m H^n \epsilon_{im}
\epsilon_{jn},~~~L^i L^j Q^k Q^l \bar{Q}_i\bar{Q}_m H^m H^n
\epsilon_{jk} \epsilon_{ln}, \nonumber \\
&~& L^i L^j Q^k Q^l \bar{Q}_i\bar{Q}_m H^m H^n \epsilon_{jn}
\epsilon_{kl},~~~L^i L^j Q^k Q^l \bar{Q}_k\bar{Q}_m H^m H^n
\epsilon_{ij} \epsilon_{ln}, \nonumber \\
&~& L^i L^j Q^k Q^l \bar{Q}_k\bar{Q}_m H^m H^n \epsilon_{il}
\epsilon_{jn},~~~L^i L^j Q^k Q^l \bar{Q}_p\bar{Q}_q H^m H^n
\epsilon_{ij} \epsilon_{km} \epsilon_{ln} \epsilon^{pq}
\nonumber \\
&~& L^i L^j Q^k Q^l \bar{Q}_p\bar{Q}_q H^m H^n \epsilon_{ik}
\epsilon_{jm} \epsilon_{ln} \epsilon^{pq},~~~
L^i L^j Q^k Q^l \bar{Q}_p\bar{Q}_q H^m H^n \epsilon_{im}
\epsilon_{jn} \epsilon_{kl} \epsilon^{pq}\} \nonumber \\
{\cal O}_{48} &=& L^i L^j d^c d^c \bar{d^c} \bar{d^c} H^k
H^l \epsilon_{ik} \epsilon_{jl} \nonumber \\
{\cal O}_{49} &=& L^i L^j d^c u^c \bar{d^c} \bar{u^c} H^k
H^l \epsilon_{ik} \epsilon_{jl} \nonumber \\
{\cal O}_{50} &=& L^i L^j d^c d^c \bar{d^c} \bar{u^c} H^k
\bar{H}_i \epsilon_{jk} \nonumber \\
{\cal O}_{51} &=& L^i L^j u^c u^c \bar{u^c} \bar{u^c} H^k
H^l \epsilon_{ik} \epsilon_{jl} \nonumber \\
{\cal O}_{52} &=& L^i L^j d^c u^c \bar{u^c} \bar{u^c} H^k
\bar{H}_i \epsilon_{jk} \nonumber \\
{\cal O}_{53} &=& L^i L^j d^c d^c \bar{u^c} \bar{u^c}
\bar{H}_i \bar{H}_j \nonumber \\
{\cal O}_{54} &=& \{L^i Q^j Q^k d^c \bar{Q}_i \bar{e^c}
H^l H^m \epsilon_{jl} \epsilon_{km},~~~L^i Q^j Q^k d^c
\bar{Q}_j \bar{e^c} H^l H^m \epsilon_{il} \epsilon_{km},
\nonumber \\
&~& L^i Q^j Q^k d^c \bar{Q}_l \bar{e^c} H^l H^m \epsilon_{im}
\epsilon_{jk},~~~L^i Q^j Q^k d^c \bar{Q}_l \bar{e^c} H^l
H^m \epsilon_{ij} \epsilon_{km}\} \nonumber \\
{\cal O}_{55} &=& \{L^i Q^j \bar{Q}_i \bar{Q}_k \bar{e^c}
\bar{u^c} H^k H^l \epsilon_{jl},~~~L^i Q^j \bar{Q}_j
\bar{Q}_k \bar{e^c} \bar{u^c} H^k H^l \epsilon_{il},
\nonumber \\
&~&L^i Q^j \bar{Q}_m \bar{Q}_n \bar{e^c} \bar{u^c} H^k H^l
\epsilon_{ik} \epsilon_{jl} \epsilon^{mn}\} \nonumber \\
{\cal O}_{56} &=& L^i Q^j d^c d^c \bar{e^c} \bar{d^c} H^k
H^l \epsilon_{ik} \epsilon_{jl} \nonumber \\
{\cal O}_{57} &=& L^i d^c \bar{Q}_j \bar{u^c} \bar{e^c}
\bar{d^c} H^j H^k \epsilon_{ik} \nonumber \\
{\cal O}_{58} &=& L^i u^c \bar{Q}_j \bar{u^c} \bar{e^c}
\bar{u^c} H^j H^k \epsilon_{ik} \nonumber \\
{\cal O}_{59} &=& L^i Q^j d^c d^c \bar{e^c} \bar{u^c}
H^k \bar{H}_i \epsilon_{jk} \nonumber \\
{\cal O}_{60} &=& L^i d^c \bar{Q}_j \bar{u^c} \bar{e^c}
\bar{u^c} H^j \bar{H}_i
\label{O6fd11}
\end{eqnarray}

\section{Renormalizable models of neutrino mass}

The classification of the effective $\Delta {\rm L} =2$ 
operators given in the previous section can be quite useful 
in building renormalizable models of neutrino mass.  
We shall describe in this section how to systematically
identify from these operators interesting neutrino mass
models.  We will see that this 
method reproduces several well--known models.  More 
interestingly, many new models of neutrino mass will 
be uncovered.  While we will not present an exhaustive 
discussion of all these new models, we will outline the 
most interesting features for neutrino mass and phenomenology
in several of these models.

\subsection{Tree--level neutrino mass models}

The operator ${\cal O}_1$ of Eq. (3) can generate small 
neutrino masses at tree level.  The simplest way to induce
${\cal O}_1$ is by the seesaw mechanism.  As shown in 
Fig. 1, ${\cal O}_1$ will result after the heavy fields 
$N_{1,3}$ are integrated out.  Here $N_1$ denotes the 
familiar $SU(2)_L$ singlet right--handed neutrinos.  
It is also possible to induce ${\cal O}_1$ using $N_3$, 
which are $SU(2)_L$ triplets and have zero hypercharge 
\cite{ma}.  

\begin{figure}[htb]
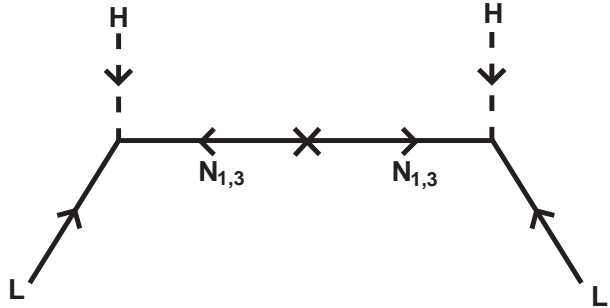

\centerline{ \DESepsf(f1.eps width 8 cm) }
\bigskip
\bigskip
\caption {\label{fig1} Tree--level neutrino mass 
generation through ${\cal O}_1$ via the seesaw mechanism. 
$N_1$~$(N_3)$ is an $SU(2)_L$ singlet (triplet) fermion 
with zero hypercharge.}
\end{figure}

\begin{figure}[htb]
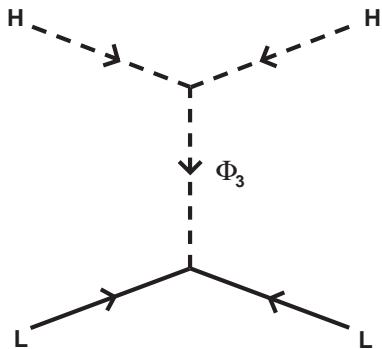

\centerline{ \DESepsf(f2.eps width 5 cm) }
\bigskip
\bigskip
\caption {\label{fig2} Type II seesaw mechanism where an
$SU(2)_L$ triplet scalar acquires an induced VEV.}
\end{figure}

In Fig. 2, we show an alternate way of inducing 
${\cal O}_1$ by the exchange of an $SU(2)_L$
triplet scalar $\Phi_3$ which carries $Y=+1$.  The 
neutral component of $\Phi_3$ will receive an induced 
vacuum expectation value (VEV) through its trilinear 
coupling with the Standard Model Higgs doublet.  This is 
sometimes referred to as the Type II seesaw mechanism 
\cite{typeII}, which can occur in the absence of 
right--handed neutrinos.  For example, in $SU(5)$ grand 
unified theories with a ${\bf 15}$--plet of scalars which 
contain the $\Phi_3$ field, the requisite trilinear scalar 
coupling will arise from the Lagrangian term ${\bf 15}~ 
{\bf \bar{5}}~{\bf \bar{5}}$, where the ${\bf 5}$--plet 
scalar fields contain the Standard Model Higgs doublet.  
The phenomenology of ${\cal O}_1$ as the source of neutrino 
mass has been studied extensively.  Here we simply note that, 
for ${\cal O}_1$ to be interesting to the current neutrino 
oscillation data, the heavy particles $N_{1,3}$ or $\Phi_3$ 
must have masses of order $10^{12}-10^{15}$ GeV.

\subsection{One--loop neutrino mass models}

The group (ii) operators listed in Eq. (4) which involve two
neutrino fields can lead to radiative neutrino masses at the 
one--loop level.  Symmetries or selection rules must forbid
the existence of the lower dimensional operator ${\cal O}_1$
for self--consistency.  We shall illustrate five examples of
one--loop neutrino mass generation.  Two of the five models
are well--known and have been thoroughly investigated in the
literature, while the other three are apparently new.

In this class of models, the induced neutrino masses are 
suppressed by a loop factor, and usually also by ratios of 
light fermion masses to the scale of new physics $\Lambda$.  
Because of the additional suppression factors, the scale 
$\Lambda$ in this class of models may be much smaller than 
the corresponding scale in the seesaw models, and may even 
be close to the electroweak scale.

\vspace*{0.1in}
\noindent{\bf Notation:}
\vspace*{0.1in}

In the models we shall present in the remainder of this 
section, heavy scalar bosons will be integrated out to 
generate the corresponding effective $\Delta {\rm L} =2$ 
operators.  We shall use the following notation to denote 
these scalars.  $\Phi$ will generically denote color
singlet scalars, $\Omega$ will denote color triplet scalars, 
and $\bar{\Omega}$ color anti--triplet scalars.  Note that 
$\bar{\Omega}$ is not the hermitian conjugate of $\Omega$.  
The $SU(2)_L$ quantum numbers of these scalar fields will 
be indicated as a subscript.  Their hypercharge quantum
numbers can be inferred from the interaction Lagrangians, 
and will not be displayed explicitly.  For example, 
$\Phi_1$ will transform as $(1,1)$ under $SU(3)_C \times 
SU(2)_L$, while $\bar{\Omega}_3$ transforms as $(\bar{3},3)$.
$SU(2)_L$ indices will be denoted by $(i,j,...)$, while the 
subscripts $(a,b,...)$ will denote generation indices.  

\vspace*{0.1in}
\noindent{\bf 1. Operator ${\cal O}_2$:}
\vspace*{0.1in}

Consider the following renormalizable Lagrangian:
\begin{equation}
{\cal L}^{{\cal O}_2} = f_{ab}L^i_a L^j_b\epsilon_{ij}
\Phi_1 + \mu \Phi_{1} \bar{H}_i \Phi_{2j}\epsilon^{ij} 
+ y_a L^i_a e^c_a\Phi_{2i} + h.c.
\label{LO2}
\end{equation}
where the Yukawa couplings $f_{ab} = -f_{ba}$ due to 
$SU(2)_L$ symmetry, $y_a$ are the (diagonal) Yukawa couplings 
of the charged leptons, and $\Phi_{2}$ is a second scalar 
doublet.  The hypercharges of $\Phi_1$ and $\Phi_2$ are 
$+1$ and $-{1 \over 2}$, respectively.  The simultaneous 
presence of the three terms in Eq. (\ref{LO2}) will
result in lepton number violation, leading to the operator
${\cal O}_2$, as depicted in Fig. 3.  Upon closing the 
$L e^c$ line (indicated by the thin solid line in Fig. 3) 
with the insertion of a Higgs field, finite neutrino masses 
will result.

\begin{figure}[htb]
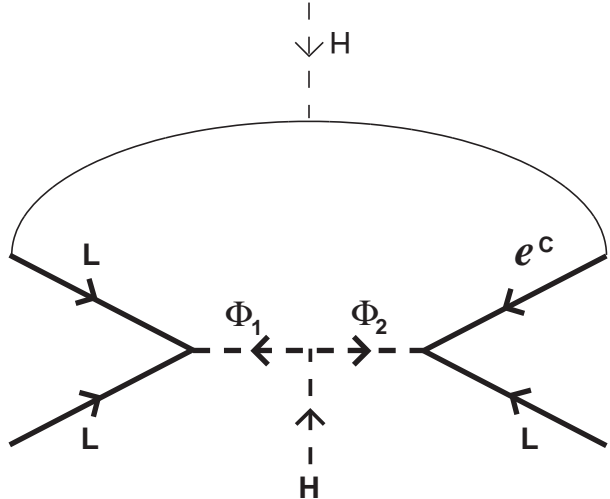

\centerline{ \DESepsf(f3.eps width 8 cm) }
\bigskip
\bigskip
\caption {\label{fig3} Diagram that induces the operator
${\cal O}_2$ as in the Zee model.}
\end{figure}

The model just described is the Zee model for neutrino
mass \cite{zee}.  The phenomenology of this model has 
been extensively studied, and we have nothing more to add 
here.

\vspace*{0.1in}
\noindent{\bf 2. Operator ${\cal O}_3$:}
\vspace*{0.1in}

The operator ${\cal O}_3$ can be induced from the 
following renormalizable Lagrangian:
\begin{equation}
{\cal L}^{{\cal O}_3} = f_{ab} L^i_a Q^j_b \epsilon_{ij}
\bar{\Omega}_{1} + g_{ab} L^i_a d^c_b \Omega_{2i} +
\mu \bar{\Omega}_{1} \Omega_2^i \bar{H}_i + h.c.
\label{LO3}
\end{equation}
where $\bar{\Omega}_1$ has hypercharge ${1 \over 3}$, 
and $\Omega_2$ has $Y= {1 \over 6}$.  It is also possible 
to write an analogous Lagrangian with ${\bar \Omega}_1$ 
replaced by $\bar{\Omega}_3$, which is a triplet of 
$SU(2)_L$, rather than a singlet.  Either of these 
Lagrangians will lead to ${\cal O}_3$ via the diagram in 
Fig. 4.  Upon closing the $Q$ line with the $d^c$ line 
(indicated by the thin solid line in Fig. 4) and 
inserting a Higgs field, this diagram will induce a 
neutrino mass at the one--loop level.  

\begin{figure}[htb]
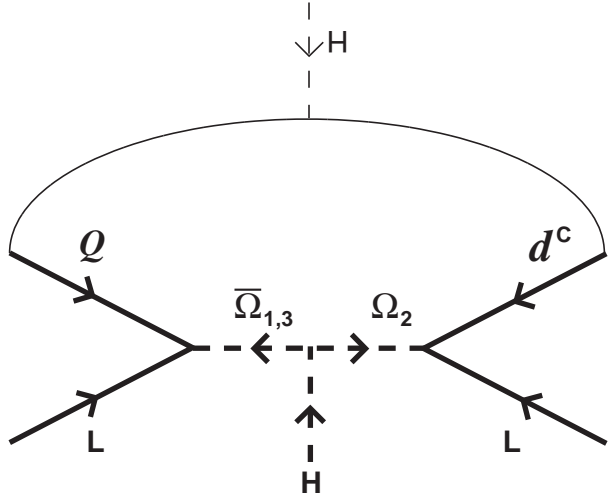

\centerline{ \DESepsf(f4.eps width 8 cm) }
\bigskip
\bigskip
\caption {\label{fig4} Diagram that induces the operator
${\cal O}_3$ through the Lagrangian in Eq. (\ref{LO3}).  
This model has a realization in supersymmetric models with 
R--parity violation.}
\end{figure}

The induced neutrino masses can be estimated from Fig. 4 
to be
\begin{equation}
m_{\nu_a} \sim {f g \over 16 \pi^2} (m_d)_a \left({\mu v
\over M^2}\right)_,
\end{equation}
where $(m_d)_a$ is the mass of the down--type quark with 
flavor index $a$, $v$ is the VEV of $H$, and $M$ is an 
average mass of the $\Omega$ fields.  Using $f = g = 10^{-3}$, 
$v = \mu = 174$ GeV and $M = 3$ TeV, we get $m_{\nu_3} 
\sim 0.06$ eV, which is in the interesting range for 
atmospheric neutrino oscillations.  Assuming minimal
flavor structure in the matrices $f$ and $g$, we also see 
that $m_{\nu_2} \sim (m_s/m_b) m_{\nu_3}$, which is in the
interesting range for solar neutrino oscillations.

A specific realization of this model (with $\bar{\Omega}_1$ 
in Fig. 4) is the supersymmetric Standard Model with 
$R$--parity violation \cite{hall}.  $\bar{\Omega}_1$ is 
identified with $\tilde{d^c}$ and $\Omega_2$ with $\tilde{Q}$ 
in this case.  The phenomenology of this specific realization 
has been well studied. We simply note that ${\cal O}_3$ has 
other realizations as well, for example, via $\bar{\Omega}_3$ 
or by using leptoquarks without supersymmetry.

\vspace*{0.1in}
\noindent{\bf 3. Operator ${\cal O}_4$:}
\vspace*{0.1in}

While it is possible to induce ${\cal O}_4$ by integrating
out scalar fields, it turns out neutrino masses will arise 
in that case either at tree level as in Fig. 2 or at 
one--loop level as in the Zee model depicted in Fig. 3.  
Consider ${\cal O}_4$ arising through the couplings 
$L L \Phi_3 + Q u^c H + H H \Phi_3^\dagger$, where $\Phi_3$ 
transforms as $(1,3,+1)$ under $SU(3)_C \times SU(2)_L 
\times U(1)_Y$.  In this case, the neutral component of 
$\Phi_3$ will acquire a tree--level VEV, and therefore 
a dimension 5 Majorana mass term for the neutrinos will 
be allowed at tree level, as in Fig. 2.  If $\Phi_3$ is 
replaced by an $SU(2)_L$ singlet $\Phi_1$ and a second 
scalar doublet $\Phi_2$ is introduced so that lepton number 
is broken, then the model becomes identical to the Zee 
model.

A more interesting way of generating ${\cal O}_4$ is to 
integrate out leptoquark gauge bosons, as shown in Fig. 5.  
The Lagrangian of the model has the form
\begin{equation}
{\cal L}^{{\cal O}_4} = \bar{Q}_i \gamma_\mu L^i G_{1}^\mu
+ \bar{u^c} \gamma_\mu L^i G_{2i}^{\mu *} + G_1^{\mu *}
G_{2 \mu}^i H^j \epsilon_{ij} + h.c.
\end{equation}
Here $G_1^\mu$ is a leptoquark gauge boson which is a
singlet of $SU(2)_L$ and has electric charge $+{2 \over 3}$.  
$G_2^\mu$ is a leptoquark gauge boson that is an $SU(2)_L$
doublet with its $I_3 = +{1 \over 2}$ member carrying an 
electric charge of $+{2 \over 3}$.  Upon spontaneous symmetry 
breaking, $G_1^\mu$ and $G_2^\mu$ will mix, leading to the 
generation of neutrino mass via the loop shown in Fig. 5.  
We have also allowed in Fig. 5 for the possibility that 
$G_1^\mu$ may be replaced by an $SU(2)_L$ triplet gauge 
boson $G_3^\mu$.

\begin{figure}[htb]
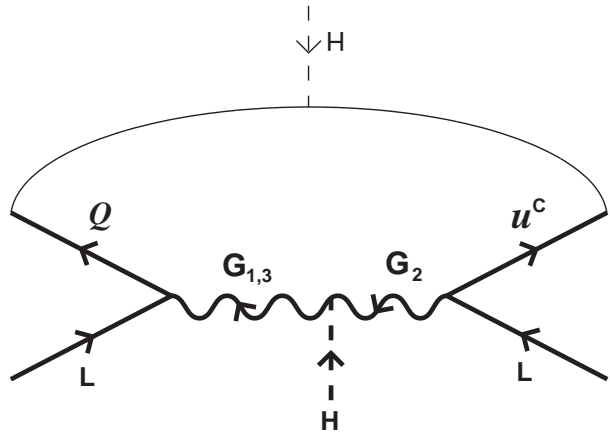

\centerline{ \DESepsf(f5.eps width 8 cm) }
\bigskip
\bigskip
\caption {\label{fig5} Leptoquark gauge bosons in $SO(10)$
GUT inducing operator ${\cal O}_4$.}
\end{figure}

In Grand Unified Theories (GUT) which unify $(Q,~L,~u^c)$ 
into a common multiplet, the diagram of Fig. 5 will arise 
naturally.  Such is the case in $SO(10)$ GUT (but not in 
$SU(5)$ GUT).  Consider the $SU(2)_L \times SU(2)_R 
\times SU(4)_C \equiv {\cal G}_{224}$ subgroup of $SO(10)$.  
The 45 gauge bosons of $SO(10)$ contain a $(1,1,15)$ and a 
$(2,2,6)$ under ${\cal G}_{224}$. $G_1$ of Fig. 5 is contained 
in $(1,1,15)$, while $G_2$ is in $(2,2,6)$.  The covariant 
derivative term for the ${\bf 16}$--dimensional Higgs 
fields will contain a piece ${\bf 45}^\dagger~ {\bf 45}~ 
{\bf 16}^\dagger~ {\bf 16}$, where ${\bf 45}$ denotes the 
gauge fields.  The Standard Model Higgs doublet $H$ may be 
contained partly in ${\bf 16}$, as in currently popular 
$SO(10)$ models \cite{so10}, ${\bf 16}$ also contains a 
Standard Model singlet component which acquires a large GUT 
scale VEV.  As a consequence, $G_1^\mu$ and $G_2^\mu$ 
will mix through this covariant derivative term.

An interesting consequence of Fig. 5 is the modification of
how the light neutrino masses scale with the quark masses in
$SO(10)$ models.  While the conventional seesaw contributions
to neutrino masses scale quadratically with the up--type 
quark masses, the new contributions arising from Fig. 5, 
given by
\begin{equation}
m_{\nu_a} \sim {g^2 \over 16 \pi^2} (m_u)_a \left({v \over M_G}
\right)_,
\end{equation}
where $g$ is the $SO(10)$ coupling constant and $M_G$ is the 
common mass of the leptoquark gauge bosons, scale linearly 
with the up--type quark masses.  Although $m_{\nu_a}$ here is 
suppressed by a loop factor, it may be the more dominant 
contribution for the lighter generations of neutrinos because 
of this linear scaling.

\vspace*{0.1in}
\noindent{\bf 4. Operator ${\cal O}_5$:}
\vspace*{0.1in}

Operator ${\cal O}_5$ is different from ${\cal O}_3$ in
that an additional $H {\bar H}$ has been added.  Notice
that the $SU(2)_L$ contraction acting on the $H {\bar H}$
is non--trivial.  Here we present a renormalizable model
which generates ${\cal O}_5$ without inducing ${\cal O}_3$.
The Lagrangian of the model is
\begin{equation}
{\cal L}^{{\cal O}_5} = f_{ab} L_a Q_b {\bar \Omega}_1 +
g_{ab} L_a d^c_b \Omega_2 + \lambda \Omega_2
{\bar \Omega}_2 {\bar H} {\bar H} + \mu {\bar \Omega}_1
{\bar \Omega}_2^\dagger H
+ h.c.
\label{LO8}
\end{equation}
We have not explicitly shown the $SU(2)_L$ contractions,
nor the hypercharges of the colored scalar fields, both
of which should be obvious from Eq. (\ref{LO8}).  This 
Lagrangian leads to operator ${\cal O}_5$ through the 
diagram of Fig. 6.  It is also possible to replace the 
${\bar \Omega}_1$ field by an $SU(2)_L$ triplet field 
${\bar \Omega}_3$, which is indicated in Fig. 6.

\begin{figure}[htb]
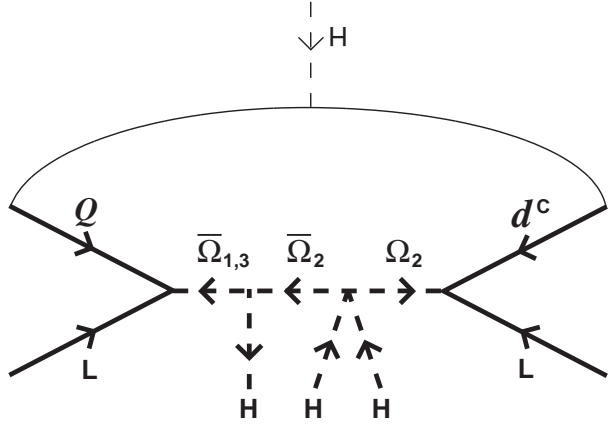

\centerline{ \DESepsf(f7.eps width 8 cm) }
\bigskip
\bigskip
\caption {\label{fig6}Feynman diagram that induces
operator ${\cal O}_5$ through Eq. (\ref{LO8}).}
\end{figure}

Upon closing the $Q d^c$ loop, we obtain a 
finite neutrino mass, which can be estimated to be
\begin{equation}
m_{\nu_a} \sim {f g \over 32 \pi^2} (m_d)_a 
\left({\lambda \mu v^3 \over M^4}\right)_,
\end{equation}
where $M$ is the assumed common mass of the colored scalars 
and $(m_d)_a$ stands for the mass of the down--type quark 
with flavor index $a$.  We can obtain interesting neutrino 
masses from Eq. (15).  For example, choose $f=g=10^{-2}$, 
$\lambda = 1$, and $\mu = M = 5$ TeV to get $m_{\nu_3} 
\sim 0.04$ eV.

\vspace*{0.1in}
\noindent{\bf 5. Neutrino mass model with a product operator:}
\vspace*{0.1in}

As noted in Sec. II, operators such as $\tilde{{\cal O}}_1 = 
L^i L^j Q^k u^c H^l H^m H^n \epsilon_{il} \epsilon_{jm}
\epsilon_{kn}$, which is a product of the Standard Model
$d=4$ operator $Q^k u^c H^n \epsilon_{kn}$ and ${\cal O}_1$, 
can lead to interesting neutrino mass models.  Here we 
illustrate this possibility with $\tilde{{\cal O}}_1$.  The 
dimension five $\Delta$L$=2$ operator will not be induced in 
this model, since we make use of a nontrivial Lorentz 
contraction.

Consider the following Lagrangian which breaks lepton number
by two units:
\begin{eqnarray}
{\cal L}^{\tilde{{\cal O}}_1} &=& f_{ab} L_a Q_b {\bar 
\Omega}_{3} + g_{ab} L_a u^c_b \Omega_2
+ \mu_1 \Omega_2^\prime \bar{\Omega}_3 \bar{H} \nonumber \\
&+& \mu_2 \Omega_2^\prime \bar{\Omega}_3^\prime H +
\mu_3 \Omega_2 {\bar \Omega}_3^\prime \bar{H} + h.c.
\end{eqnarray}
Here $(\Omega_2, \Omega_2^\prime)$ are $SU(2)_L$ doublet
leptoquark scalars with $Y=(7/6,~1/6)$ respectively, while
$(\bar{\Omega}_3, \bar{\Omega}_3^\prime)$ are $SU(2)_L$
triplet leptoquarks  with $Y = (1/3,~-2/3)$ respectively.
We have suppressed $SU(2)_L$ indices for simplicity.

\begin{figure}[htb]
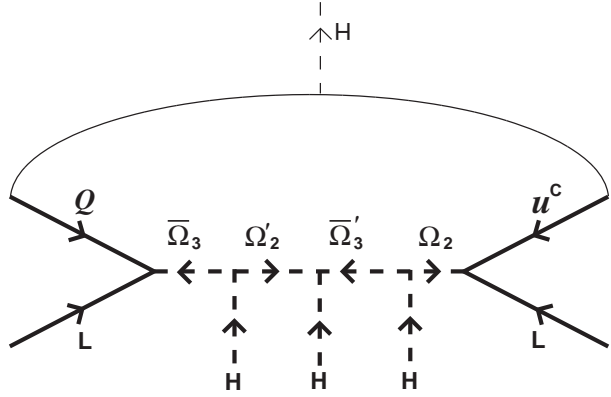

\centerline{ \DESepsf(f6.eps width 8 cm) }
\bigskip
\bigskip
\caption {\label{fig7}Scalar exchange contributions 
inducing operator $\tilde{{\cal O}}_1$ (see text).}

\end{figure}

The diagram shown in Fig. 7 will generate
$\tilde{{\cal O}}_1$ with this Lagrangian.  Upon closing the 
$Q$ and $u^c$ lines, as shown by the light solid line in 
Fig. 7, neutrino masses will be induced as one--loop 
radiative corrections.  We can estimate the induced neutrino 
masses to be
\begin{equation}
m_{\nu_a} \sim {f g \over 48 \pi^2}
(m_u)_a \left({\mu_1 \mu_2 \mu_3 v^3 \over M^6}\right)_,
\label{mO7}
\end{equation}
where we have ignored any generation dependence in the
coupling matrices $f$ and $g$, and assumed a common mass $M$ 
for the leptoquark scalars, taken to be much heavier than 
the weak scale.  In Eq. (\ref{mO7}), $(m_u)_a$ denotes the
mass of the up--type quark with flavor index $a$.  Take, as 
an example, $f=g=10^{-2}$, $\mu_1=\mu_2 = \mu_3 = M = 10$ TeV.  
In this case, $m_{\nu_3} \sim 0.2$ eV, which is in the 
interesting range for atmospheric neutrino oscillations.  
Clearly, other choices of parameters are possible with 
interesting neutrino phenomenology even with the scale of new 
physics being relatively low.

Another feature worth mentioning is that the lowest
dimensional $\Delta {\rm L} = 2$ operator that arises in this
model is $\tilde{{\cal O}}_1$ as shown in Fig. 7.  We can assign
lepton number of $-1$ to both $\Omega_2$ and $\bar{\Omega}_3$ 
that have Yukawa couplings to leptons.  It is the mixing term 
$\Omega_2 \bar{\Omega}_3$ that violates L by two units.  
However, this mixing can occur in the model only after 
inserting three $H$ fields, as in the figure.  As a result, 
no lower dimensional $\Delta {\rm L} = 2$ operators are 
induced.

If we replace the color anti--triplet scalars in Fig. 7,
which are triplets of $SU(2)_L$, by color anti--triplets
that are $SU(2)_L$ singlets, the contribution to 
$\tilde{{\cal O}}_1$ will vanish, since the effective scalar 
mixing term $\Omega_2^\dagger {\bar \Omega}_1^\dagger H H H$ 
is identically zero due to $SU(2)_L$ symmetry.

\subsection{Two--loop neutrino mass models}

Let us now turn to renormalizable models wherein neutrino
masses are induced as two--loop radiative corrections.  A
generic feature of this class of models is that, for 
interesting neutrino mass phenomenology, the scale of new 
physics tends to be much lower than the corresponding scale 
in one--loop mass generation models.  In addition to an 
extra loop suppression factor, we will see that the induced 
neutrino masses in this class of models scale quadratically 
with charged fermion masses, as opposed to the typical linear 
scaling in the one--loop models.

\vspace*{0.1in}
\noindent{\bf 1. Operator ${\cal O}_{9}$:}
\vspace*{0.1in}

Neutrino mass models induced via ${\cal O}_{9}$ have been
well studied \cite{zee1,babu}.  The relevant Lagrangian is
\begin{equation}
{\cal L}^{{\cal O}_{9}} = f_{ab} L_a L_b \Phi_1 + g_{ab}
e^c_a e^c_b \Phi_1^{--} + \mu \Phi_1 \Phi_1 \Phi_1^{--} 
+ h.c.
\label{LO12}
\end{equation}
where $\Phi_1$ is a singly charged $SU(2)_L$ singlet scalar, 
while $\Phi_1^{--}$ is a doubly charged singlet.  Operator
${\cal O}_{9}$ is induced through the diagram of Fig. 8. 
(Such a two--loop diagram for neutrino mass generation was 
first discussed in Ref. \cite{cheng}.)

\begin{figure}[htb]
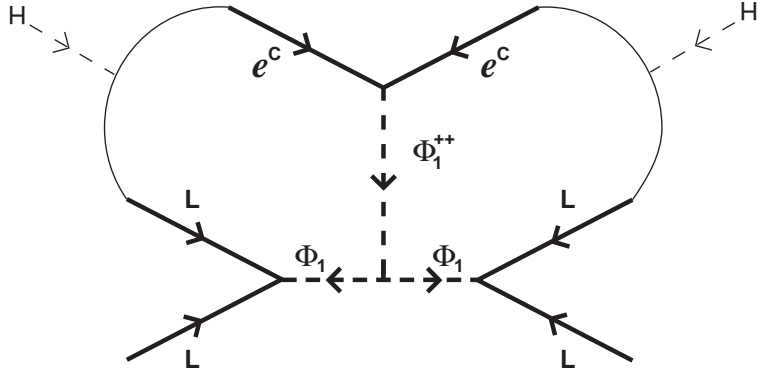

\centerline{ \DESepsf(f8.eps width 10 cm) }
\bigskip
\bigskip
\caption {\label{fig8} Operator ${\cal O}_{9}$ and the
generation of two--loop neutrino masses in the model of 
Eq. (\ref{LO12}).}
\end{figure}

Upon closing the $L e^c$ loops on both sides, 
indicated by the light solid lines in Fig. 8, we see  
that the induced neutrino mass matrix has a structure 
given by
\begin{equation}
m_{\nu} = {1 \over (16 \pi^2)^2}\left(f M_\ell g M_{\ell}
f^T\right) \left({\mu \over M^2}\right)_.
\label{mO12}
\end{equation}
Here $M_{\ell}$ stands for the diagonal charged lepton
mass matrix.  An interesting feature of this matrix
structure is that, owing to $SU(2)_L$ symmetry,
$f_{ab} = -f_{ba}$, which means that, for three
generations of neutrinos, Det($m_{\nu}) = 0$.  This
indicates that one of the neutrinos will remain massless
at the two--loop level \cite{babu}, but it will acquire a 
mass at the three--loop level.

To see the numerical magnitude of the induced neutrino
masses, let us set $f=g=0.09$, $\mu = M = 3$ TeV.  The
heaviest neutrino will then have a mass $m_{\nu_3}
\sim 0.03$ eV.

It should be noted that ${\cal O}_{9}$ can also arise
if $\Phi_1$ in Fig. 8 is replaced by $\Phi_3$, an
$SU(2)_L$ triplet scalar.

\vspace*{0.1in}
\noindent{\bf 2. Operator ${\cal O}_{10}$:}
\vspace*{0.1in}

We shall present two renormalizable models that induce
${\cal O}_{10}$.  The effective operators are depicted
for the two models in Figs. 9 and 10.  The renormalizable
Lagrangians for the two models can be readily written 
down from these figures.  They are
\begin{eqnarray}
{\cal L}_1^{{\cal O}_{10}} &=& f_{ab} L_a L_b \Phi_1 +
g_{ab} L_a Q_b {\bar \Omega}_1
+ h_{ab} e^c_a d^c_b \Omega_1 + \mu \Omega_1
{\bar \Omega}_1 \Phi_1 + h.c. \nonumber \\
{\cal L}_2^{{\cal O}_{10}} &=& f_{ab} L_a L_b \Phi_1 +
g_{ab} L_a d^c_b \Omega_2
+ h_{ab} e^c_a Q_b {\bar \Omega}_2 + \mu \Omega_2
{\bar \Omega}_2 \Phi_1 + h.c.
\end{eqnarray}
Here the two Lagrangians are meant to be taken 
separately, and not simultaneously.  As shown in Figs. 9
and 10 in light solid lines, upon annihilating the $L$ 
and $e^c$ fields as well as the $Q$ and $d^c$ fields, 
two--loop neutrino masses will result.  The estimate of 
the masses is analogous to that in Eq. (\ref{mO12}), 
with one difference, namely, one of the charged lepton 
mass matrices must be replaced by the down--type quark 
mass matrix.

\begin{figure}[htb]
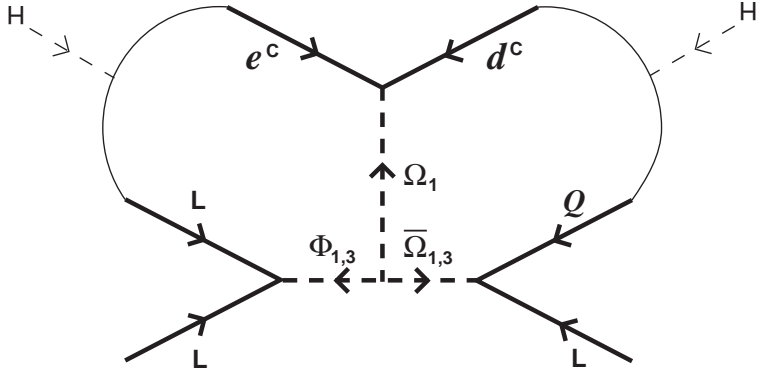

\centerline{ \DESepsf(f9.eps width 10 cm) }
\bigskip
\bigskip
\caption {\label{fig9}Diagram inducing ${\cal O}_{10}$
through ${\cal L}_1^{{\cal O}_{10}}$.}
\end{figure}

\begin{figure}[htb]
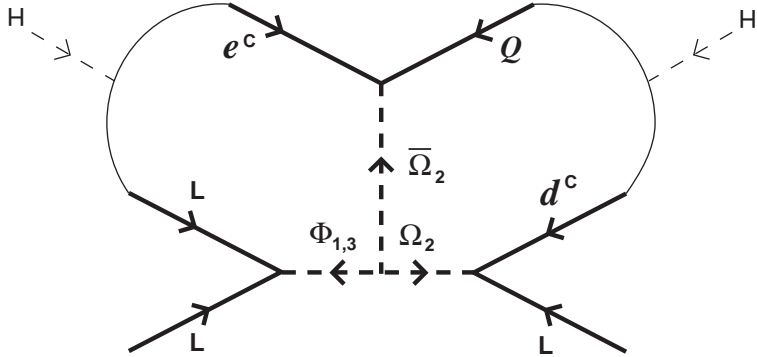

\centerline{ \DESepsf(f10.eps width 10 cm) }
\bigskip
\bigskip
\caption {\label{fig10}Diagram inducing ${\cal O}_{10}$
through ${\cal L}_2^{{\cal O}_{10}}$.}
\end{figure}

\indent It is possible to substitute an $SU(2)_L$ 
triplet scalar $\Phi_3$ for the $SU(2)_L$ singlet 
$\Phi_1$ in these models.  This is also indicated in 
Figs. 9 and 10.  In this case, a symmetry must be 
used to prevent the coupling $\Phi_3 \bar{H} \bar{H}$ 
so that the neutral component of $\Phi_3$ does not 
acquire an induced VEV at tree level, or else the
dimension 5 term of Fig. 2 will contribute more 
dominantly to neutrino masses.  Clearly, both models 
will lead to interesting neutrino masses, e.g., with 
$m_{\nu_3} \sim 0.03$ eV, if the scale of new physics 
is around a few TeV.

\vspace*{0.1in}
\noindent{\bf 3. Operator ${\cal O}_{11}$:}
\vspace*{0.1in}

Operator ${\cal O}_{11}$ contains the fields
$(LL QQ d^c d^c)$.  We shall present three models for
inducing this effective operator.  The three models 
correspond to the following different contractions:  
(1) $(LL)(QQ)(d^c d^c)$, (2) $(LQ)(LQ)(d^c d^c)$, and
(3) $(Ld^c)(Ld^c)(QQ)$.  Consider the $(LL)(QQ)(d^c d^c)$
contraction first.  The required couplings in the
renormalizable model are shown in Fig. 11.  The relevant
Lagrangian for this model is
\begin{equation}
{\cal L}_1^{{\cal O}_{11}} = f_{ab} L_a L_b \Phi_1 +
g_{ab} Q_a Q_b \Omega_1 +
h_{ab} d^c_a d^c_b {\bar \Omega}_1 + \mu \Phi_1
{\bar \Omega}_1 \Omega_1 + h.c.
\label{LO14a}
\end{equation}

As shown in Fig. 11, the scalar $\Phi_1$ may be replaced
by an $SU(2)_L$ triplet scalar $\Phi_3$.  In this case,
the coupling $\Phi_3 \bar{H} \bar{H}$ must be prevented 
by a symmetry so that the neutral component of $\Phi_3$ 
does not acquire an induced VEV at tree level, or else 
the dimension 5 operator of Fig. 2 will be the more 
dominant effective neutrino mass operator.  An example 
of such a symmetry is a $Z_4$, under which $\Phi_3 
\rightarrow - \Phi_3$, $L \rightarrow i L$, $e^c 
\rightarrow i e^c$, with other fields being neutral.  
The cubic scalar coupling term in Eq. (\ref{LO14a}) 
breaks this $Z_4$ symmetry, either softly or 
spontaneously.

\begin{figure}[htb]
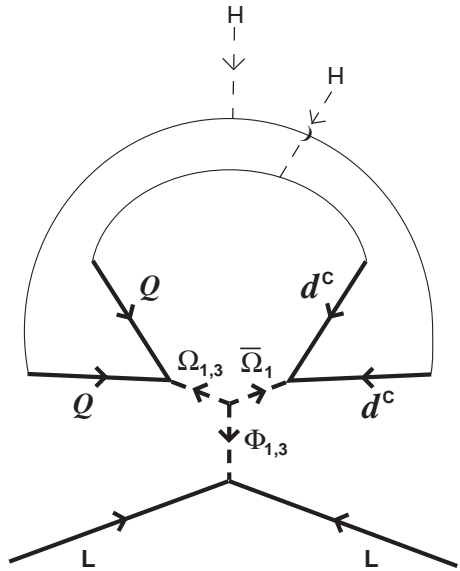

\centerline{ \DESepsf(f11.eps width 6 cm) }
\bigskip
\bigskip
\caption {\label{fig11} An example of two--loop 
radiative generation of neutrino masses from the 
operator ${\cal O}_{11}$.}
\end{figure}

Upon closing the quark lines in Fig. 11, the 
effective scalar coupling $\Phi_3 \bar{H} \bar{H}$ will 
be generated so that the neutral member of 
$\Phi_3$ will get a VEV.  This VEV will be suppressed 
by two--loop factors and two powers of down--type quark 
masses, and will scale as $v^2/M$, where $M$ is the 
assumed common mass of the heavy scalars.  The neutrino 
masses that are induced will be approximately 
\begin{equation}
m_{\nu_a} \sim {fgh \over (16 \pi^2)^2} 
\left({\mu (m_d)_a^2 \over M^2}\right)_.
\end{equation}
The numerical estimate of the masses parallels that of 
Eq. (\ref{mO12}).

The case of using an $SU(2)_L$ singlet scalar $\Phi_1$, 
as in Eq. (\ref{LO14a}), has an interesting consequence.  
The effective $d=9$ operator will be of the form 
$(\nu e)(ud) (d^cd^c)$.  In order to convert it into a
neutrino mass term, we can contract one of the $d^c$ 
fields with the $d$ field, but we will be left with 
$(\nu e)(u d^c)$.  Further exchange of a $W$ boson can 
convert the $e$ into a $\nu$ and the $u$ into a $d$.  
This is shown in Fig. 12.  Note that only the 
longitudinal component of $W$ contributes to this 
diagram, and we have indicated it as $H^-$ in Fig. 12.  
To see this, it is convenient to work in the Landau gauge, 
where the $W$ boson propagator is purely transversal.  In 
this gauge, it is clear that the amplitude for the 
transition of an off--shell vector boson into a scalar 
boson must vanish. It can, however, arise through the 
longitudinal component of the vector boson.  (Such higher 
loop diagrams for neutrino masses have been considered
in Ref. \cite{babuma}.)

\begin{figure}[htb]
\centerline{ \DESepsf(f12.eps width 6 cm) }
\bigskip
\bigskip
\caption {\label{fig12} A three--loop diagram that
induces neutrino masses from the operator ${\cal O}_{11}$ 
through the Lagrangian of Eq. (\ref{LO14a}).}
\end{figure}

\indent In addition to an extra loop suppression factor, 
the three--loop diagram of Fig. 12 is suppressed by 
couplings of the longitudinal $W$ to charged fermions.  
We estimate the induced neutrino masses to be
\begin{equation}
m_{\nu} \sim { f g h \over (16 \pi^2)^3} \left(
{\mu m_\ell^2 m_d^2 \over M_W^2 M^2}\right)_,
\end{equation}
where $m_\ell$ and $m_d$ denote the respective charged
lepton and down--type quark masses, and $M_W$ is the mass 
of the $W$ boson.  With this estimate, we see that, for 
neutrino masses to be in the interesting range, the 
masses of the scalars should be of the same order as 
the weak scale.

In Fig. 13, we display the renormalizable model where
the fermion fields in ${\cal O}_{11}$ are contracted as 
$(LQ)(LQ)(d^cd^c)$.  The Lagrangian terms of this model 
are
\begin{equation}
{\cal L}_2^{{\cal O}_{11}} = f_{ab} L_a Q_b
{\bar \Omega}_1 + g_{ab} d^c_a d^c_b
{\bar \Omega}^\prime_1 + \mu {\bar \Omega}_1
{\bar \Omega}_1 {\bar \Omega}^\prime_1 + h.c.
\label{LO14b}
\end{equation}
We can also replace the ${\bar \Omega}_1$ field by
${\bar \Omega}_3$, an $SU(2)_L$ triplet.

\begin{figure}[htb]
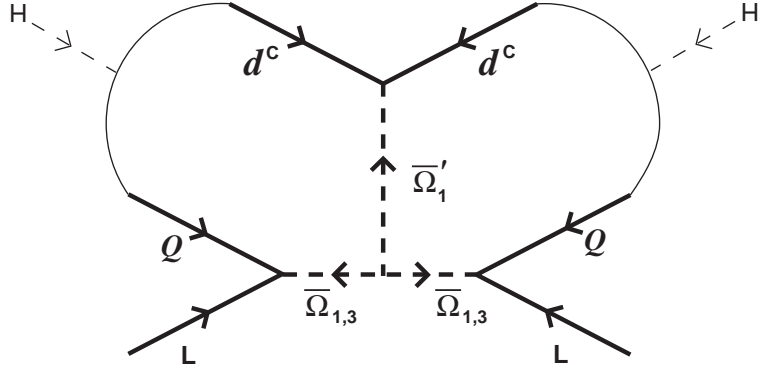

\centerline{ \DESepsf(f13.eps width 10 cm) }
\bigskip
\bigskip
\caption {\label{fig13} Diagram that induces the operator
${\cal O}_{11}$ through Eq. (\ref{LO14b}).}
\end{figure}

Fig. 14 shows the fermion field contraction
$(Ld^c)(Ld^c)(QQ)$ that induces ${\cal O}_{11}$.  The
Lagrangian for this case is
\begin{equation}
{\cal L}_3^{{\cal O}_{11}} = f_{ab} L_a d^c_b \Omega_2
+ g_{ab} Q_a Q_b \Omega_1 + \mu \Omega_2 \Omega_2 \Omega_1
+ h.c.
\label{LO14c}
\end{equation}

\begin{figure}[htb]
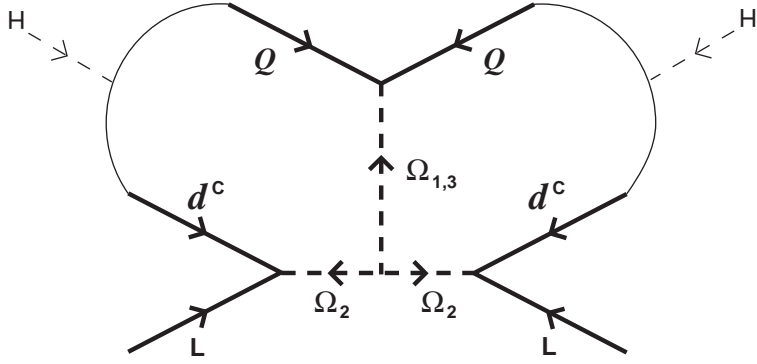

\centerline{ \DESepsf(f14.eps width 10 cm) }
\bigskip
\bigskip
\caption {\label{fig14} Two--loop neutrino mass generation 
from ${\cal O}_{11}$ through the renormalizable Lagrangian 
in Eq. (\ref{LO14c}).}
\end{figure}

In both Figs. 13 and 14, the induced neutrino masses are 
of the same order, given approximately by 
\begin{equation}
m_{\nu_a} \sim {f^2 g \over (16 \pi^2)^2} (m_d)^2_a 
\left({\mu \over M^2}\right)_.
\end{equation}

\vspace*{0.1in}
\noindent{\bf 4. Operator ${\cal O}_{12}$:}
\vspace*{0.1in}

${\cal O}_{12}$ can be obtained in a way very similar 
to ${\cal O}_{11}$.  The neutrino mass generation 
mechanism is shown in Fig. 15.  Note that this 
diagram is very similar to Fig. 11, with the $d^c$ 
fields in Fig. 11 replaced by the $u^c$ fields here.  
We simply write down the Lagrangian for this model:
\begin{equation}
{\cal L}^{{\cal O}_{12}} = f_{ab} L_a L_b \Phi_1 +
g_{ab} Q_a Q_b \Omega_1 + h_{ab} u^c_a u^c_b
{\bar \Omega}_1 + \mu {\bar \Omega}_1 \Omega_1
\Phi_1^\dagger + h.c.
\label{LO15}
\end{equation}

\begin{figure}[htb]
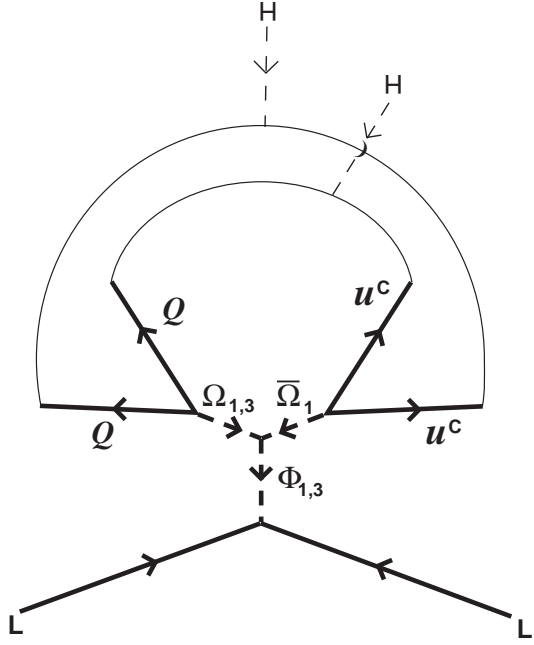

\centerline{ \DESepsf(f15.eps width 7 cm) }
\bigskip
\bigskip
\caption {\label{fig15} Diagram inducing ${\cal O}_{12}$
through Eq. (\ref{LO15}).}
\end{figure}

\noindent Compared to Fig. 11, the only difference in
the neutrino mass estimate is that here it will be 
proportional to $m_u^2$, rather than $m_d^2$.

\vspace*{0.1in}
\noindent{\bf 5. Operator ${\cal O}_{21}$:}
\vspace*{0.1in}

${\cal O}_{21}$ is a dimension 11 operator.  We shall
illustrate various ways of building renormalizable 
models of neutrino mass with this operator.  The 
field content of ${\cal O}_{21}$ is $(LL Le^c Qu^c HH)$.  
As in the previous examples, we can contract the fermion
fields in different ways.  Three specific models are 
obtained by the contractions (1) $(LL)(e^c u^c)(QL)$,
(2) $(LQ)(Le^c)(Lu^c)$, and (3) $(LL)(Lu^c)(Qe^c)$.  
The Lagrangians for these three choices are as follows:
\begin{eqnarray}
{\cal L}_1^{{\cal O}_{21}} &=& f_{ab} L_a L_b \Phi_1 +
g_{ab} e^c_a u^c_b \Omega_1
+ h_{ab} L_a Q_b {\bar \Omega}_1 + [\Phi_1 \Omega_1
{\bar \Omega}_1 \bar{H} \bar{H}]
+ h.c. \nonumber \\
{\cal L}_2^{{\cal O}_{21}} &=& f_{ab} L_a Q_b
{\bar \Omega}_1  + g_{ab} L_a e^c_b \Phi_2
+ h_{ab} L_a u^c_b \Omega_2 + [\Phi_2 \Omega_2
{\bar \Omega}_1 \bar{H} \bar{H}] 
+ h.c. \nonumber \\
{\cal L}_3^{{\cal O}_{21}} &=& f_{ab} L_a L_b \Phi_1 +
g_{ab} L_a u^c_b \Omega_2  + h_{ab} Q_a e^c_b {\bar
\Omega}_2 + [\Phi_1 {\bar \Omega}_2 \Omega_2  \bar{H} 
\bar{H}] + h.c.
\label{LO24}
\end{eqnarray}
Here we have not explicitly written down the renormalizable 
terms that result in the scalar mixings.  These scalar 
mixing terms can be read off from Figs. 16, 17 and 18.

\begin{figure}[htb]
\centerline{ \DESepsf(f16.eps width 8 cm) }
\bigskip
\bigskip
\caption {\label{fig16} Generation of the $d=11$ 
operator ${\cal O}_{21}$ through the Lagrangian 
${\cal L}_1^{{\cal O}_{21}}$ of Eq. (\ref{LO24}).}
\end{figure}

\begin{figure}[htb]
\centerline{ \DESepsf(f17.eps width 12 cm) }
\bigskip
\bigskip
\caption {\label{fig17} Generation of the $d=11$ 
operator ${\cal O}_{21}$ through the Lagrangian 
${\cal L}_2^{{\cal O}_{21}}$ of Eq. (\ref{LO24}).}
\end{figure}

\begin{figure}[htb]
\centerline{ \DESepsf(f18.eps width 12 cm) }
\bigskip
\bigskip
\caption {\label{fig18} Generation of the $d=11$ 
operator ${\cal O}_{21}$ through the Lagrangian 
${\cal L}_3^{{\cal O}_{21}}$ of Eq. (\ref{LO24}).}
\end{figure}

Neutrino masses arise in all three models upon 
closing the $Le^c$ lines and the $Q u^c$ lines.  The 
order-of-magnitude estimates for the masses in the 
case of Figs. 17 and 18 are
\begin{equation}
m_{\nu_a} \sim \left({fgh \over (16 \pi^2)^2}\right)
\left({(m_\ell)_a (m_u)_a \mu^3 v^2 \over M^6} \right)_.
\end{equation}
Here we have denoted by $\mu$ all cubic scalar couplings
entering the diagrams.  $(m_\ell)_a$ and $(m_u)_a$ denote, 
respectively, the masses of the charged leptons and the 
up--type quarks with flavor index $a$.  If we choose 
$f=g=h=0.15$ and $\mu = M = 3$ TeV, we obtain 
$m_{\nu_3} \sim 0.05$ eV, 
in the interesting range for atmospheric neutrino 
oscillations.  The same estimate will apply to Fig. 16 if 
the $\Phi_3$ scalar is used.  On the other hand, if 
$\Phi_1$ is used, there is an additional loop suppression, 
along with a suppression factor of $m_\ell^2/M_W^2$ as in 
Fig. 12.

\section{Neutrinoless double beta decays and neutrino mass}

In many of the $\Delta {\rm L} = 2$ effective operators, 
there is a tree--level contribution to neutrinoless double
beta ($\beta\beta 0\nu$) decay amplitudes, while neutrino
masses arise only as two--loop radiative corrections.\footnote{
Gauge models where this phenomenon occurs have been noted 
in specific contexts \cite{beta}.}  Such models will have 
an exciting phenomenological consequence, namely, that 
$\beta\beta 0\nu$ decays might be observable in the 
current round of experiments, while the induced neutrino 
masses are quite consistent with the atmospheric and solar 
neutrino data.

The effective operators that mediate $\beta\beta 0\nu$
processes will have the form $[uu \bar{d} \bar{d} e e]$.
Operators of this form can be easily identified from our 
list.  The $d=9$ operators from Eq. (7) that have this 
form are:

\begin{equation}
{\cal O}_{\beta\beta 0\nu}^{d=9} = \{{\cal O}_{11}^{(ii)},
~ {\cal O}_{12}^{(i)},~{\cal O}_{14}^{(ii)},~{\cal O}_{19},
~ {\cal O}_{20}\}_.
\end{equation}
Here the superscript ${(ii)}$ stands for the second entry
in the respective operator, and so on.  From the set of
$d=11$ operators in Eq. (8), the following operators will
induce $\beta\beta 0\nu$ decays at tree level:
\begin{eqnarray}
{\cal O}_{\beta\beta 0\nu}^{d=11} &=& \{ {\cal O}_{24}^{(i)},
~{\cal O}_{28}^{(i),(iii)},~ {\cal O}_{32}^{(i)},
~{\cal O}_{36},~{\cal O}_{37},~ {\cal O}_{38}, \nonumber \\
&~& {\cal O}_{47}^{(i),(iv)},~{\cal O}_{53},
~ {\cal O}_{54}^{(i),(iv)},~ {\cal O}_{55}^{(i)},~
{\cal O}_{59},~ {\cal O}_{60}\}_.
\end{eqnarray}

In order to estimate the strength of these $\beta\beta
0\nu$ operators, we compare their amplitudes to that
arising from light Majorana neutrino exchanges.  This latter 
amplitude is given by $A_{\beta \beta 0\nu} \sim G_F^2 m_\nu 
\left<{1 \over q^2} \right>$, where $ \left <{1 \over q^2}
\right>$ is the average of the inverse squared Fermi momentum 
and is approximately equal to $1/(100$ MeV)$^2$.  The current
experimental constraint on light Majorana neutrino masses 
is $m_\nu < 0.3$ eV, derived from the nonobservation of 
$\beta \beta 0\nu$ decays \cite{klapdor}.  Hence, 
$A_{\beta \beta 0\nu} < 10^{-18} ~{\rm GeV}^{-5}$.  Consider 
now the amplitude arising from ${\cal O}_{11}^{(ii)}$.  
Since this operator has dimension 9, it carries a 
suppression factor of $\Lambda^{-5}$, where $\Lambda$ 
denotes the scale of new physics above which 
the effective description becomes invalid.  Comparing the 
two amplitudes, we see that, if $\Lambda \sim 10^{3.6}$ 
GeV $\sim 4$ TeV, the induced $\beta\beta 0\nu$ decay 
amplitude will be near the current experimental limit.  
From the diagram of Fig. 11 (with $\Phi_3$ field rather
than $\Phi_1$ field), we also see that $m_\nu \sim 
(1/16 \pi^2)^2 m_d^2/\Lambda \sim 3 \times 10^{-4}$ eV, 
which is extremely tiny and well consistent with solar 
and atmospheric neutrino data.  (Here we considered only 
the first generation couplings and assumed the relevant 
$f \sim g \sim h \sim 1$.) Similar estimates apply to the 
other $d=9$ operators as well.  For the $d=11$ operators,
there is an additional suppression factor of
$(v/\Lambda)^2$ due to the presence of two additional
Higgs fields.  The scale $\Lambda$ will have to be 
somewhat smaller than 4 TeV in these cases for
$\beta\beta 0\nu$ decays to be observable in the near 
future.  The induced neutrino masses will also have this 
same additional suppression factor.  Thus, it is possible 
to have negligible neutrino masses while having sizable 
$\beta\beta 0\nu$ processes.

Of course, our estimates here have been rather crude.  
We have ignored important differences in nuclear matrix
elements.  However, we think that the above 
order-of-magnitude estimates suffice to demonstrate the 
potential importance of these operators.  It will be 
interesting to investigate in more detail the rates for 
$\beta\beta 0\nu$ processes in this class of models.

\section{Conclusions}

We have presented a classification of effective
$\Delta {\rm L} = 2$ operators for the Standard Model
that may be relevant for generating small neutrino 
Majorana masses.  The lowest dimensional ($d=5$) such 
operators are provided by the seesaw mechanism.  If 
these $d=5$ operators are absent from the underlying 
renormalizable gauge theory due to selection rules, 
higher dimensional operators will become relevant.  
These higher dimensional $\Delta {\rm L} = 2$ operators 
will induce neutrino masses through radiative corrections.  
The scale of new physics, $\Lambda$, can be as low as a 
few TeV in this class of models.  

We have presented a list of all $\Delta {\rm L} = 2$ 
operators through $d=11$ which contain no derivatives or 
gauge boson fields.  We can readily construct from this 
list various renormalizable models for neutrino masses.  
We are able to identify several of the well--known 
radiative neutrino mass models as specific realizations of 
some of these effective operators.  Furthermore, we can 
identify a large class of new models where neutrino masses 
arise as one--loop, two--loop or even three--loop radiative 
corrections.  We have given several examples of these new 
models, and have outlined their main features for neutrino 
mass phenomenology.  Of special interest is a class of 
operators of dimension 9 and 11 which contribute directly 
to neutrinoless double beta decays, while generating only 
very tiny neutrino masses at the two--loop level.  Models 
of this class have the exciting prospect that neutrinoless 
double beta decays will be observable in the ongoing round 
of experiments, while being fully consistent with solar and 
atmospheric neutrino oscillations.  The operators presented
here will have interesting consequences for cosmology as 
well.  For example, the observed baryon asymmetry in the 
universe may have been generated by electroweak sphaleron 
processes which converted a primoridal lepton asymmetry 
into baryon excess.  The $\Delta$L$=2$ operators listed 
here will be constrained if we require that the observed 
baryon asymmetry is produced correctly by this 
mechanism \cite{fy}.  We plan to revisit our models in a 
future publication and analyze this and other issues such 
as the expectation for flavor changing neutral current 
processes in the leptonic as well as in the hadronic 
sectors.

\newpage
\centerline{\bf Acknowledgements}
\bigskip
The work of KB is supported in part by DOE Grant 
\# DE-FG03-98ER-41076, a grant from the Research Corporation,
DOE Grant \# DE-FG02-01ER4864 and by the OSU Environmental 
Institute.  CNL is supported in part by the U.S. Department 
of Energy under grant DE-FG02-84ER40163 and in part by the 
University of Melbourne through its Visiting Scholars program.  
He is grateful for the hospitality extended to him by S. Nandi, 
P. Tong, X.C. Xie, K. Hamanaka, M. Babu and other members of 
the Physics Department at Oklahoma State University where this 
work was initiated.  He would also like to acknowledge the warm 
hospitality from the High Energy Theory groups in the School 
of Physics at the University of Melbourne and in the Institute 
of Physics at the Academia Sinica in Taipei, Taiwan, where 
part of this work was carried out.

\end{document}